\documentclass[aps,prb,twocolumn,groupedaddress]{revtex4}
\usepackage{amssymb}
\usepackage{graphicx}
\usepackage{epsfig}
\usepackage{float}

\newcommand{\be}{\begin{equation}}
\newcommand{\ee}{\end{equation}}
\begin{document}

\title{Optical Properties and Correlation Effects in
Na$_x$CoO$_2$}
\author{M. D. Johannes, I. I. Mazin, D. J. Singh}
\affiliation{Code 6391, Naval Research Laboratory, Washington, D.C. 20375}

\begin{abstract} We have calculated the optical spectra of Na$_{x}$CoO$_2$ for $x$=0.3, 0.5, and 0.7
within the LDA. We compare our results to available experimental data and show that the important
features and trends are reproduced well, but there is a nearly uniform shift of peak positions and poor
agreement in intensities.  We show, through application of a simple model, that these differences can
be attributed to overhybridization between Co and O orbitals and spin fluctuations which renormalize
the bandwidth.  Applying the LDA+U procedure shifts the optical peaks further from their experimental
locations, indicating that this method of incorporating correlation effects is ill-suited for the case
Na$_{x}$CoO$_2$. \end{abstract}

\maketitle

The layered cobaltate Na$_x$CoO$_2$ is the subject of considerable interest not only because of a
possibly unconventional superconducting state \cite{KTHS+03}, but also an unusual phase diagram that
encompasses two very different metallic states, and at least one insulating region \cite{MLF+04}, and
unusual magnetotransport and thermoelectric behavior.  \cite{YWNSR+03,ITYS97,SYL+04} For $x <$ 0.5,
the system behaves as a paramagnetic metal with Pauli-like susceptibility and superconducts when
hydrated in the narrow region around $x$=0.3.  At $x$=0.5, a charge ordered insulating state, which is
likely anti-ferromagnetic, occurs.  When $x$ is increased above 0.5, the metallicity returns, but the
susceptibility becomes Curie-Weiss like, \cite{JSJHB+03} and a spin-density wave appears \cite{JSHI+03}
around $x$=0.75.  It has been recently reported \cite{JHJY+04} that, at low temperatures, a second
insulating state may set in at $x$=0.25.

LDA calculations show \cite{MDJ+} a band structure that evolves smoothly with $x$, in contrast to
the sharply different regions described above.  The experimental phase diagram is not well
reproduced, and a weakly ferromagnetic, metallic ground state is found incorrectly for all Na
concentrations.  Neglect of strong correlation effects by the LDA is the most obvious culprit, but
their precise role in Na$_{x}$CoO$_{2}$ is not understood. Circumstantial evidence indicates that
it is substantial, with most estimates of the Hubbard $U$ quite large in comparison with the
$t_{2g}$ bandwidth. However, the system does not behave characteristically as a doped Mott-Hubbard
insulator, particularly around the superconducting composition where susceptibility measurements
indicate a lack of local magnetic moments on Co $d$ ions.\cite{RJBCS+03,FCC+03} Also important is
the observation that LDA calculations in systems with strong Hubbard correlations routinely
underestimate the tendency to magnetism (the larger the difference, $U-I$, between the Hubbard
repulsion and Hund rule coupling, the greater the underestimation) \cite{AGP+03}, and overestimate
the antiferromagnetic superexchange $J.$ In the LDA, Na$_{x}$CoO$_{2}$ is magnetic \cite{DJS00},
with antiferromagnetic instabilities that are smaller than ferromagnetic. This requires
effects beyond LDA to suppress magnetism and casts doubt on frustration as a
mechanism for suppression of magnetism. Because of this, despite the large $U/W$ ratio, spin
fluctuations may offer a more plausible reason for the deviations from the LDA.

In this context, it is important to verify to what extent LDA bands correctly describe the one-electron
excitations and fermiology of Na$_{x}$CoO$_{2}.$ Quantum oscillation probes would be ideal to
investigate the Fermi surface (FS), particularly the predicted small pockets \cite{DJS00,MDJ+b}, but so
far no such studies have been reported.  Several photoemission reports are available
\cite{MZH+03,H-Y03}, but because the electronic structure is expected to be sensitive to both surface
states and surface quality, generalization of these results to the details of bulk electronic
properties is questionable.  Infrared optics has a penetration depth of $\sim$ $c/\omega _{p}$ (where
$\omega _{pl}$ is the plasma frequency), much longer than that of photoemission ($\sim$ 10 $\AA$), and
may therefore provide the most reliable probe of the electronic structure available to date. Several of
these experiments have been reported \cite{SLMO04,JHJY+04,NLW+,CBAVB+, GCLD+04}. However, since optical
absorption is an integrated property, involving a variety of interband transitions, it is usually
difficult, if even possible, to interpret the results in terms of the electronic structure without full
first principles calculations.

Here we report such calculations.  By comparing our calculated LDA spectra with optical
experiments reported for various Na contents, we are able to assign the three main peaks to
specific interband transitions and thereby examine in detail the dependence of these peak
positions and strengths on bandwidths and energies.  The peak shifts one expects to see in
Mott-Hubbard type compounds are not realized in this system, even for the insulating compounds.  
Furthermore, application of the LDA+U methodology exaggerates the systematic discrepancies between
the LDA and measured spectra. We employ a simple model to illustrate that overestimation of Co-O
hybridization due to overly extended d-orbitals would produce precisely the kind of peak
shifts we see in our LDA spectra.  Spin fluctuations predicted early on \cite{DJS03} and later
observed \cite{ATB+03} at some values of $x$, could also affect the spectra by renormalizing the
bandwidth of the $t_{2g}$ complex.  This is also consistent with the differences we observe in our
transitions compared with experiment.

\section{Calculated Optical Conductivity}

Using the Wien2k \cite{Wien2k} full-potential augmented plane wave + local orbital code (APW+lo), we
calculated the optical conductivity for Na$_x$CoO$ _2$ at three Na contents: $x$=0.3,0.5 and 0.7. To
vary the Na content, we used the virtual crystal procedure of occuping each site with an ion of
fractional charge. For both $c$ and $a$ lattice constants, we used experimental data \cite{MLF+04} but
relaxed the apical O height separately at each $x$. The APW sphere radii were the same for all
calculations: 2.0 for Na, 1.85 for Co and 1.55 for O. $Rk_{max}$ (essentially a measure of the
plane-wave cut-off) was set to 7.0, giving a basis set of 833 APW's and 60 local orbitals. Our density
was well converged using 480 $k$-points in the irreducible Brillouin zone (BZ).

\begin{figure}[tbp] \includegraphics[width= .95\linewidth]{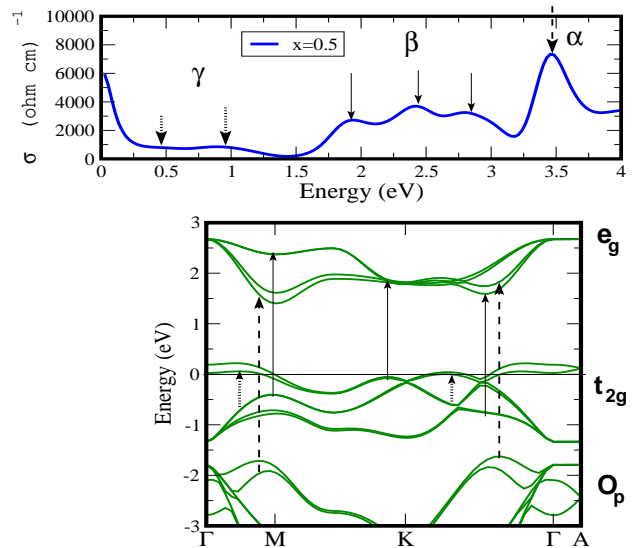}
\caption{\textit{top panel} The LDA optical conductivity for
Na$_{0.5}$CoO$ _2 $ with arrows delineating the three main peaks:
$\protect\gamma$ (lined arrows), $\protect\beta$ (solid arrows) and
$\protect\alpha$ (dotted arrows) A Gaussian broadening of 0.06 $eV$ was used.
\textit{bottom panel} Interband transitions corresponding to peaks in the
optical conductivity.} \label{trans} \end{figure}

We first calculated the imaginary part of the dielectric constant, $\epsilon''(\omega)$ in the
random phase approximation (RPA) using the standard code of the Wien2k package. The real part of
the optical conductivity, $\sigma'(\omega)$, can then be obtained by $\sigma'(\omega)$ $=$
$\frac{\omega}{4\pi}\epsilon''(\omega)$. The resulting spectra exhibit three main peaks, which we
label $\gamma$, $\beta$, and $\alpha$, according to the notation of Wang \textit{et al}
\cite{NLW+}.  We can identify the specific interband transitions giving rise to these peaks using
a band by band decomposition of $\epsilon''(\omega)$.  Each of $\gamma$, $\beta$, and $\alpha$
have distinct origins. The $\gamma$-peak corresponds to transitions between different bands of the
metallic $t_{2g}$ complex, the $\beta$-peak to $t_{2g}$-$e_g$ transitions, and the $\alpha$-peak
to transitions from fully occupied O-$ p$ states to unoccupied $e_g$ states.  Note that the
$\gamma$ and $\beta$ peaks exist despite the dipole selection rule which forbids $d$-$d$
transitions, because of O hybridization throughout both $d$-band complexes. The $t_{2g}$ bands are
split into an $a_{1g}$ and two $e_g'$ bands by the trigonal crystal field \cite{DJS00}, and all
$\gamma$ transitions are from a lower $e_g'$-like state to an upper $e_g'$ or $a_{1g}$-like
state.  The e$_g'$-e$_g'$ transitions are wholly due to small hole FS's near the $K$ points of the
BZ and contribute to the lowest visible peak in the calculated spectrum (the $\gamma$ subpeak near
0.4 eV in Fig. 1).  However, there are allowed $e_g'$-$a_{1g}$ transitions in the same energy
range that contribute roughly equally to the peak intensity and therefore, the existence of these
hole pockets cannot be inferred from the existence of the peak.  The calculated $\beta$ peak has
several distinguishable subpeaks corresponding to transitions from different parts of the BZ. In
Fig \ref{trans}, the peaks of the spectrum for Na$_{0.5}$CoO$_2$ are shown and representative
transitions are indicated along high symmetry points of the corresponding band structure.

\section{Comparison with Experiment}

\begin{figure}[]
\includegraphics[width = 0.95\linewidth,angle=270]{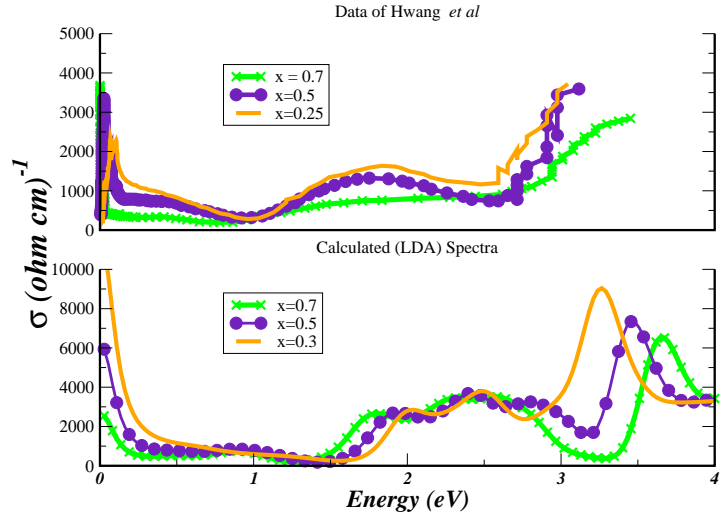}
\caption{(color online) A comparison of experimental (top panel)  and calculated (lower panel)  optical 
spectra.  Note that the 
lowest calculated Na concentration is 0.3, slightly different from the experimental value of 0.25.}
\label{comp}
\end{figure}

The three ($\gamma $, $\beta $, $ \alpha $) conductivity peaks are clearly recognizable in all
available optical data. The reported positions and strengths of these peaks, however, vary
somewhat between experiments, even for identical dopings \cite {NLW+,GCLD+04,JHJY+04}. For
consistency of comparison, we use mainly the report of Hwang \textit{et al} \cite{JHJY+04}, where
spectra for several compounds of different Na content are available. Overall, the LDA spectra are
seen to be in good qualitative, but poor quantitative agreement with experiment (see Fig.
\ref{comp}), a situation we will attribute to LDA overhybridization, and/or spin fluctuations in
the next section. But first, we discuss the doping dependencies and general shape of the spectra.  
These are well reproduced by calculation. There are two clear trends followed by both the LDA and
measured $\sigma _{1}(\omega )$ as electron count is increased (Na added). First, the $\alpha $
and $\beta $ peaks shift away from each other and become broader. Second, the $\gamma $ peak
sharpens and moves slightly lower in energy. Both effects have their origin in an upward shift of
Co-derived bands as the Co $d$ band filling is increased. The greater energy difference between
the O-$p$ states and the shifted $e_{g}$ band determines the $\alpha $ peak position, and a
reduction in hybridization between Co and O orbitals, now further separated in energy, weakens the
$t_{2g}$-$e_{g}^{\prime }$ crystal field splitting, causing the downward $\beta $-peak shift.
Since the $\gamma $ -peak is due to transitions within the $t_{2g}$ band complex, its sharpening
and shift down are both attributable to a slightly narrowed $t_{2g}$ bandwidth caused by decreased
O mixing. Our calculated in-plane plasma frequency at $x$=0.7, $\omega _{p}$ = 1.39 $eV$, agrees
well with values extracted from experiment \cite{NLW+,GCLD+04}: $\omega _{p}$ = 1.48 $eV$ and
$\omega _{p}$ = 1.17 $eV$. The c-axis plasma frequency is calculated to be 1.48 $eV$, nearly the
same as the in-plane value, reflecting the lack of anisotropy at higher Na levels \cite{VCAnote}.
As Na is removed, the in-plane plasma frequency and the electronic anisotropy increase until, at
$x$=3, the in-plane $\omega _{p}$ is 2.86 $eV$ and the c-axis $\omega _{p}$ is 0.337 $ eV $,
approximately an order of magnitude apart.  The increase in Drude weight and concurrent
enhancement of anisotropy with electron removal are in good agreement with experimental
observations \cite{JHJY+04, CBAVB+}.

Despite these instances of good agreement, some dramatic differences between the spectra are
obvious. The experimental peak intensities are almost a factor of two smaller and approximately
0.5 eV lower in energy than our calculated ones when a Gaussian broadening of 0.06 $eV$ (this
value reproduces all three reported peak widths well) is used. Experimentally, the strong dip just
below the $\alpha$ peak deepens from $x$=0.25 to $x$=0.5, but then becomes shallow again at
$x$=0.7, whereas the LDA gap deepens and widens monotonically with increasing Na content.  It has
been previously suggested both $\beta$ and $\alpha$ peaks could be identified with transitions
across $t_{2g}$-$e_g$ energy gap and that spin splitting of $t_{2g}$ states could account for
differences between observed peak positions and calculated paramagnetic band positions
\cite{NLW+}. Though no observation of static magnetism has been made for any $x<0.75$, the LDA is
known to predict an FM ground state for all $x$, suggesting that magnetic fluctuations may be
present in the system \cite {DJS00,DJS03}. Direct observations of FM spin fluctuations \cite
{ATB+03} and a spin density wave \cite{JSHI+03} at $x$=0.75, and c-axis AFM ordering \cite{SPB+04}
at $x$=0.85 further support the idea that magnetism may play a role in optics.  However, since
regular optics does not allow for spin-flip transition, spin splitting may shift optical
transitions only if one of the two electronic states involved in an transtion is split and the
other is not. This is not the case for either $\gamma$ or $\beta$ peaks. There is some effect on
the position of the $\alpha$ peak, because of oxygen admixture, but it is merely 0.15 eV even at
$x$=0.3. Thus, the observed shifts of the absorption peaks compared to LDA, especially the two
high-energy peaks, $\alpha$ and $\beta$, cannot be due to magnetic effects.

\section{Correlations Effects and Band Renormalizations}

The small $t_{2g}$ and $e_{g}$ bandwidths of Na$_{x}$CoO$_{2}$ suggest that correlation effects in
Co-d may be important. Our renormalized atom calculations for Hubbard $U$ on Co yield $U\gtrsim
$3.7 eV, while the $t_{2g}$ bandwidth $W$ is 1.5 eV, so that the Hubbard ratio (accounting for
degeneracy) is $U/W\sqrt{3}\gtrsim 1.5$. Spin-unrestricted band structure calculations yield a
half-metallic ferromagnetic ground state, contrary to experiment.  All this suggests that
conventional LDA calculations should be taken with a grain of salt in this compound, and it is
quite likely that the actual experimental electronic structure is considerably renormalized
compared to the LDA results. Indeed, angle-resolved photoemission spectroscopy (ARPES)
measurements \cite{H-Y03,MZH+03} have detected a heavy quasiparticle band near the Fermi energy
with a total width of $\sim $ 70 $meV$, much smaller than the LDA bandwidth. This is an
interesting finding, but surface sensitivity should be kept in mind. Indeed, differences between
surface electronic structure, as measured accurately by ARPES, and bulk electronic structure have
been established in other oxides, including cuprate \cite{WEP+92}, ruthenate \cite{APM03}, and
manganite \cite{J-P+98,BNIIM+01} compounds. Optical experiments, with their longer penetration
depth have the potential to tell us more about the specific shortcomings of LDA calculations in
sodium cobaltate.

Let us first recall the typical inadequacies of LDA calculations in systems with localized
electrons. The best known case is exemplified by Mott-Hubbard insulators. Here the missing physics
is mainly the Coulomb repulsion between electrons localized at the same site. Typical examples are
$f-$electron compounds, high-T$_{c}$ cuprates, or 3$d$ oxides. This group is characterized by (1)  
underestimation of propensity toward magnetism (up to the level of total loss of a magnetic ground
state, as in cuprates) (2) underestimation of band gaps between occupied and empty bands and (3)
overestimation of the superexchange antiferromagnetic interaction (this increases with hopping and
decreases with the band gap; the former is overestimated and the latter underestimated in LDA). As
a leading correction to the LDA, the LDA+U method has been very successful in the prototypical
systems listed above. For these, the LDA+U method reproduces the missing Mott-Hubbard effects,
such as \textquotedblleft sinking\textquotedblright\ of the occupied bands relative to the
unoccupied bands (with corresponding shifts of interband transitions to higher energies).

Inspection of the LDA bandstructure of Na$_{x}$CoO$_{2}$ in conjunction with experimental data
indicates that this physics is not applicable here. The tendency toward magnetism is
\textit{over}estimated, and so is the band gap between the $t_{2g}$ and $e_{g}$ bands. The LDA ground
state is ferromagnetic, and not antiferromagnetic.  Not surprisingly, LDA+U calculations
\cite{JKK-L03,PZWL+04} only worsen the situation. The $ t_{2g}-e_{g}$ distance, already 0.5 eV too
large, increases further, and the tendency towards magnetism becomes even stronger. Additionally, for
more subtle reasons, the two $e_{g}^{\prime }$ subbands of the $t_{2g}$ manifold are shifted down with
respect to the $a_{1g}$ band. This in turn shifts the $\gamma$ peak, already too high in energy, up
even further. We conclude that the \textit{LDA+U approach is not appropriate for
Na}$_{x}$\textit{CoO}$_{2}.$

Another possible manifestation of electron-electron correlations we is found in $^{3}$He and in
CrO$_{2}$ \cite{IIM99}. Here collective excitations (magnons or paramagnons) play the same role as
phonons in the sense of \textquotedblleft dressing\textquotedblright\ quasiparticles and increasing
their mass \cite{NFB96}. As a result, the optical spectra more or less proportionally squeeze toward
the low frequency. Softening of magnetic excitations near a quantum critical point leads to large spin
fluctuations and suppression (full or partial) of magnetic ordering \cite{IIM03b,TM85}. A comparison of
the experimental and calculated properties of Na$_{x}$CoO$_{2}$ indicates the presence of such effects.  
Thus, \textit{quantum criticality is likely to be an important reason for deviation from the LDA in
Na}$ _{x}$\textit{CoO}$_{2}.$

Finally, one should not forget about a very prosaic shortcoming of LDA: the fact that it includes a
spurious self-interaction which leads to overextended 3$d$ orbitals, and thus overestimated
hybridization with ligand orbitals. This effect becomes stronger for more localized $d$-orbitals. For
the narrow $d$-bands of Na$_{x}$CoO$ _{2}$, this will be operative to some extent.

We shall now try to assess qualitatively the ramification of these two
effects. The first, \textquotedblleft dressing\textquotedblright\ of
one-electron excitations is qualitatively similar to the electron-phonon
coupling induced renormalization, except it occurs in a larger energy
range. It is reasonable to expect it to affect the whole $t_{2g}$ band or
a large part of it, leading to overall narrowing of all three $t_{2g}$
bands. This, in turn should shift both components of the $ \gamma $
transition, the $e_{g}^{\prime }-e_{g}^{\prime }$ one and the $
e_{g}^{\prime }-a_{1g}$ one to lower energies, without much change in
their intensity (the extra spectral weight is transferred to high
energies, as in the electron-phonon coupling). It is unlikely that, as
sometimes assumed, the top of the $e_{g}^{\prime }$ band will be shifted
down with respect to the $ a_{1g} $ band. First, dressing of the
quasiparticles makes them heavier, but normally does not shift different
states near the Fermi energy with respect to each other. Second, this
would not only eliminate the $e_{g}^{\prime }-e_{g}^{\prime }$
transitions, but would also shift the $e_{g}^{\prime }-a_{1g}$ to higher
energy, opposite to what is required by the experiment. It is also
unlikely that such \textquotedblleft dressing\textquotedblright\ would
affect the unoccupied $e_{g}$ band, which is too far away from the Fermi
level.

Let us now estimate the possible effects of contraction of the Co $d$ -orbitals. To this
end we use the simplest possible tight binding model, namely one where only nearest neighbor Co-O
$pd\sigma $ and $pd\pi $ hoppings (we will these $t_{\sigma }^{2}$ and $t_{\pi }^{2}$ respectively) are
included, and the energy of the O $p$ level is taken to be sufficently far removed from the Co $d$
level to be integrated out.  Both assumptions are extremely crude, for direct Co-Co and especially O-O
overlaps are not small, nor is the $p-d$ energy separation small, but the model nonetheless provides a
useful qualitative framework. The resulting 5x5 Hamiltonian has a 2x2 $e_{g}$ subblock with diagonal
elements proportional to $t_{\sigma }^{2}.$ Because there is no path connecting an $e_{g}$ orbital on
one site to any neighboring $e_{g}$ orbital via oxygen, there is no dispersion in this subblock.  The
3x3 $t_{2g}$ subblock has noticeable dispersion controlled by $t_{\pi }^{2},$ with the overall upward
shift of the same order. Finally there is a dispersive off-diagonal subblock, corresponding to
O-assisted hopping between neighboring $t_{2g}$ and $e_{g}$ orbitals. The scale of this block is set by
the product $t_{\sigma }t_{\pi }.$ Note that in a cubic structure, such $t_{2g}$-$e_{g}$ hopping
through an intermediary O is impossible.

We now make a further simplifying assumption, that $t_{\sigma }\gg t_{\pi }.$ We can then neglect the
ligand field on the $t_{2g}$ orbitals (proportional to $t_{\pi }^{2},$), leaving only the ligand field
on the $e_{g}$ orbitals (proportional to $t_{\sigma }^{2}$).  Since the $\beta $ peak is due to
$t_{2g}-e_{g}$ transitions, its position reflects a crystal field splitting of the Co $d$-states which
results from a combination of electrostatic and ligand field effects.  Using a linearized muffin-tin
orbital (LMTO) calculation with the Co-O hybridization suppressed, we found that the crystal field
splitting reduces from $\approx 3.4$ eV to $ \approx 1$ eV. That is, the electrostatic crystal field is
1 eV, and the ligand field is 2.4 eV. The latter is probably overestimated because the LDA
overhybridizes the Co $d$ orbitals. Specifically, if $t_{\sigma }$ were 12\% smaller than it's LDA
value ($\sqrt{2.4/1.9}=1.12),$ the the $\beta $ peak would shift down by 0.5 eV, in agreement with the
experiment. Furthermore, the intensity of this peak would also be reduced, because the optical matrix
elements in a two-band tight binding model scale with the off-diagonal elements of the Hamiltonian
(this follows from the tight-binding defininition of dipole matrix elements as $\left\langle
A_{1}|\partial H/\partial \mathbf{k}|A_{2}\right\rangle ,$ where $A$ are the eigenvectors), and these
scale as $t_{\sigma }t_{\pi }.$ This is in accord with overestimation of the $\beta $ peak intensity
and lends further support to our assumption that the LDA overestimates the $pd\sigma $ hopping
amplitudes in Na$_{x}$CoO$_{2}$, probably by some 10\%.

\section{Conclusions.} We have calculated the LDA optical conductivity spectra of Na$_x$CoO$_2$ and
compared it with experimental data for several different Na contents.  There is good qualitative
agreement in terms of the number of peaks and their behavior with changing $x$, but exact peak heights
and positions are not well-reproduced.  Though the underlying reasons for this are likely correlation
effects neglected by the LDA, we argue that using LDA+U as a remedy is inappropriate for this compound.
Spin fluctuation driven renormalization of the $t_{2g}$ band complex and overextension of Co
$d$-orbitals are shown to affect the optical spectrum in a manner consistent with the discrepancies
between calculation and experiment.  These cause narrowing of the $t_{2g}$ band and reduction of
crystal field splitting respectively, bringing LDA optical peaks into good agreement with experimental
reports.

We would like to thank C. Ambrosch-Draxl, B. Batlogg, A. Boothroyd, R. Jin, W. Koshibae, and I.  Terasaki for valuable input and discussions.  M.D.J.
is supported by a National Research Council Associateship.  Work at the Naval Research Laboratory is supported by the Office of Naval Research.

\end{document}